\documentstyle[sprocl,epsfig]{article}
\def\be{\begin{equation}}
\def\ee{\end{equation}}
\def\bea{\begin{eqnarray}}
\def\eea{\end{eqnarray}}
\newcommand{\xbj}{x}
\begin{document}
\title{Resolved Photons and BFKL-type Signatures 
 in Deep Inelastic Scattering 
 \footnote{To be published in Proceedings of the {\em Madrid low x Workshop},
 Miraflores de la Sierra, 1997 }
 }
\author{ H.~Jung}
\address{University of Lund, Department of Physics, Sweden}
\maketitle\abstracts{
The concept of resolved virtual photons 
in addition to direct deep inelastic
$ep$ scattering is used to 
simulate the $2+1$ jet - rate and the forward jet cross section,
which cannot be described by direct LO/NLO processes.
With 
standard DGLAP evolution of the parton densities of the virtual photon and 
 the proton the HERA data can be described,
 provided a sufficiently large scale for the
hard scattering matrix element is used.}
%%%%%%%%%%%%%%%%%%%%%%
\section{Introduction}
%%%%%%%%%%%%%%%%%%%%%%
At HERA the measurement of the
inclusive forward jet production  \cite{Mueller_fjets1,Mueller_fjets2}
in the proton direction in deep inelastic events at low $x$
and large $Q^2$ 
is believed to be one  of 
the most sensitive measurements for BFKL dynamics. 
The measured cross section \cite{H1_fjets_data} 
cannot be explained
by standard deep inelastic scattering models and
even next - to - leading - order calculations (NLO, order $\alpha_s^2$) 
predict a too small cross section \cite{NLO_fjets}.
\par
The  measured \cite{H1_2+1jets_data}  
fractional $2+1$ jet rate $R_{2+1} = \frac{\sigma^{2+1}}{\sigma^{tot}}$,
where two high $p_T$ jets come from the hard QCD scattering and the
"+1" stands for the remnant jet, seems to be a factor of $\sim 2$
 larger than  
predictions from standard deep inelastic scattering (DIS) models.
\par
However both measurements are rather well described by the 
Color - Dipole - Model (CDM) in its implementation in
ARIADNE~\cite{CDM}. 
%Comparing ARIADNE with standard DIS fixed order
%matrix element calculations, it has been shown in \cite{Rathsman_CDM_MEPS} 
%that the boson - gluon - fusion part agrees rather well, whereas the 
%QCD - Compton part is much larger in ARIADNE, than expected from 
%any matrix element calculation.
\par
In this article we study to what extent 
the contribution of virtual resolved photons 
in addition to the standard DIS matrix element calculations (labeled as
$direct$ contributions) can reproduce the data.
 The studies are performed with the
RAPGAP~\cite{RAPGAP206} Monte Carlo event generator which 
for the direct processes 
agrees rather well with the LEPTO~\cite{Ingelman_LEPTO65} Monte Carlo program. 
%%%%%%%%%%%%%%%%%%%%%%%%%%%%%%%%%%%%
\section{Resolved Photons in DIS}

%%%%%%%%%%%%%%%%%%%%%%%%%%%%%%%%%%%
Resolved photon processes play an important role in photo-production,
when high $p_T$ jets are observed.
Any internal
structure of the proton as well as of the photon can be resolved as long
as the scale $\mu^2$ 
of the hard subprocess, which is of the order of $p_T^2$,
is larger than the inverse radius of the proton 
$1/R^2_p \sim \Lambda_{QCD}^2$
 and the photon 
$1/R^2_{\gamma} \sim Q^2$.
\par
Resolved photon processes in DIS are calculated here using the
Equivalent Photon Approximation, giving the flux of virtual transverse 
polarized photons~\cite[and references therein]{RAPGAP206,RAPGAP} together
with the on shell matrix elements for the partonic subprocesses.
 The structure
of the virtual photon is given by parametrisations of the 
parton densities~\cite{GRS,Sasgam,Drees_Godbole}
$x_{\gamma} f_{\gamma}(x_{\gamma},\mu^2,Q^2)$,
 now depending on the scale $\mu^2$ and the photon virtuality $Q^2$.
  Direct deep inelastic scattering and
resolved virtual photon processes are simulated using the RAPGAP 
 Monte Carlo event generator~\cite{RAPGAP206}.
The hard subprocesses for resolved photons implemented in RAPGAP are:
$ gg \rightarrow q \bar{q}$,
$ g g \rightarrow gg$, 
$ q g \rightarrow q g $, 
$ q \bar{q} \rightarrow g g $, 
$ q \bar{q} \rightarrow q \bar{q}$ and 
$ q q \rightarrow q q $. 
\par
In DIS, resolved photon processes \cite{Chyla_res_gamma,GRS}
can play a role when the scale $\mu^2$
 of the hard subprocess is
larger than the virtuality of the photon $Q^2$:
 $\mu^2 > Q^2$. A typical NLO diagram like $\gamma^* g \to q \bar{q} g$ 
occurs also in the
resolved photon picture, via the subprocess 
$\bar{q} g \to \bar{q} g$ with the photon splitting $\gamma^* \to q \bar{q}$.
When including resolved photon processes in a NLO 
 calculation  special care
has to be taken to avoid double counting \cite{Kramer_res_gamma}. In leading
order there is no double counting when adding  resolved photons to direct 
processes since they have different final states (three parton and two parton).
%%%%%%%%%%%%%%%%%%%%%%%%%%%%%%%%%%%%
\subsection{Scales, $\alpha_s$ and Parton Distribution Functions}
%%%%%%%%%%%%%%%%%%%%%%%%%%%%%%%%%%%%
In leading order $\alpha_s$ processes
the renormalisation scale $\mu_R$ 
%(which enters in $\alpha_s (\mu_R^2)$) 
and factorization scale $\mu_F$ 
%($xf(x,\mu_F^2)$) 
are not well defined, and any choice of 
$\mu_R^2 = \mu_F^2 =Q^2,\;p_T^2,\; Q^2 + p_T^2$, and multiples
thereof, is reasonable. 
There are essentially two competing effects:
a larger scale gives smaller $\alpha_s(\mu^2)$, but
a larger parton density $xf(x,\mu^2)$ at fixed $x$.
The net effect depends on the
details of the interaction.
\par
The effect of different 
choices of the scale 
$\mu^2 = Q^2 + 4 \cdot p_T^2$, $\mu^2 = Q^2 +  p_T^2$,
$\mu^2 = 4 \cdot p_T^2$, or $\mu^2 = p_T^2$
is discussed in the following.
\par
 We use the GRV 94 HO (DIS) parametrisation of the proton
parton densities~\cite{GRVa,GRVb} together with the 
 SaSgam~2D~\cite{Sasgam} parametrisation for the partons in the
virtual photon. Other choices like the GRV LO parton density of
the real photon together with the 
virtual photon suppression factor of Drees - Godbole~\cite{Drees_Godbole} 
give similar results.
%%%%%%%%%%%%%%%%%%%%%%%%%%%
\section{Forward Jets}
%%%%%%%%%%%%%%%%%%%%%%%%%%%
We study the HERA data on high $p_T$ ``forward" jet production
 at low $x$
and large $Q^2$ 
in the context of resolved virtual photons. 
\begin{figure}[htb]
\begin{center}
\epsfig{figure=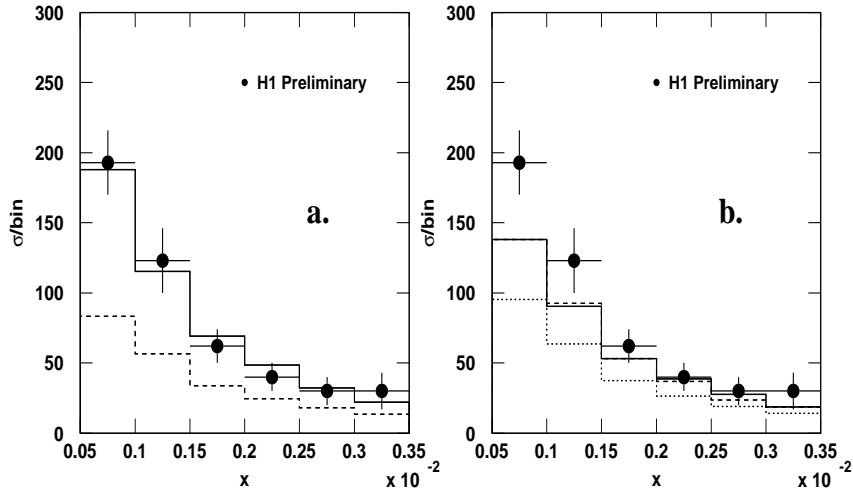,
width=11cm,height=6.5cm}
\end{center}
\caption{
The forward jet cross section as a function of $\xbj$ using the
cuts of the H1 collaboration specified in the text. The data points
are preliminary H1 data \protect\cite{H1_fjets_data}. 
In $1a._)$ the dashed line is the RAPGAP  prediction
for direct DIS processes and the solid line is the RAPGAP 
prediction for the sum of direct and resolved contributions
(using $\mu^2=Q^2 + 4 p_T^2$, GRV HO for the proton and
SaSgam 2 DIS for the virtual photon).
In $1b._)$ the solid line is the RAPGAP  prediction
for the sum of direct and resolved contributions 
with $\mu^2=Q^2 + p_T^2$, 
the dashed line with $\mu^2 = 4 p_T^2$ and the dotted line with
$\mu^2 = p_T^2$. The difference between solid and dashed lines is hardly
visible.
\label{fjet_res+dir}
}
\end{figure}
 The cuts are taken from
the preliminary H1 analysis~\cite{H1_fjets_data}: 
$y>0.1$, $E'_e > 11$ GeV, $ 160^o < \theta _{e} < 173^o$,
$7^o < \theta _{jet} < 20^o$, 
$E_{jet} > 28.7$  GeV, and $p_T^{jet} > 3.5$ GeV
in the laboratory frame.
 The observed forward jet is required to have sufficiently large
 $E_{jet}$   and $p_T^{jet}$  
 in order to rely on well measured parton distributions
 in the proton. A $p_T$ of the jet is required
to be of the same order as $Q^2$ ($0.5 < p^2_T/Q^2 < 2$)
  to suppress the contribution from  direct LO processes
which are governed by DGLAP evolution.
 The event topology is a small $\xbj$
deep inelastic event with an energetic high $p_T$ jet in the forward
(proton) direction.   
In Fig.~\ref{fjet_res+dir}$a._)$ preliminary data of the 
H1 collaboration  \cite{H1_fjets_data} are 
shown together with the RAPGAP~\cite{RAPGAP206} 
Monte Carlo prediction for $standard$ DIS processes ($direct$) 
and the sum of direct and resolved processes, using 
the scale $\mu^2=Q^2 + 4 p_T^2$. We note that the 
dominant contribution of resolved photon processes to the forward jet cross
section comes from
$qg \to qg$.
\par
 With a  scale $\mu^2 = Q^2 + 4 p_{T}^2$ (with $p_{T}$ defined
in the parton parton cms)  a perfect description of the data
(Fig.~\ref{fjet_res+dir}$a._)$) is achieved.
The effect of different choices of the scale $\mu^2$ is shown in 
Fig.~\ref{fjet_res+dir}$b._)$.
We observe no difference using $\mu^2 = Q^2 + p_T^2$ or 
$\mu^2 = 4 p_T^2$. 
In general a less hard scale results in a smaller cross section and 
 worse agreement with the data at small $\xbj$ values.
 Using
$\mu^2 = p_{T}^2$ the predicted cross section falls well below the
measurement.
This is because the condition $\mu^2=p_T^2 > Q^2$ together with
$0.5 < Q^2/p_T^2 < 2$ leaves only little space for resolved photon 
contributions. However
with the concept of  resolved virtual photons we obtain a improved 
description of the data for all choices of the scale $\mu^2$.
%%%%%%%%%%%%%%%%%%%%%%%%%%%%%%%%
\section{$2+1$ jet rates in DIS}
%%%%%%%%%%%%%%%%%%%%%%%%%%%%%%%%
Dijets ($2+1$ - jets) are 
searched for  with a cone jet 
algorithm by requiring a $p_T^{jets}> 5$ GeV in the hadronic center of 
mass frame, $Q^2>5 $ GeV$^2$, 
$y>0.05$, $E' > 11$ GeV and $156^o < \theta_e < 173^o$
 according to the preliminary H1 - analysis~\cite{H1_2+1jets_data}.
Since in that analysis $p_T^2 > Q^2$ in the region $Q^2< 25$ GeV$^2$, 
resolved photon interactions can be expected to play a significant role.
\begin{figure}[htb]
\begin{center}
\epsfig{figure=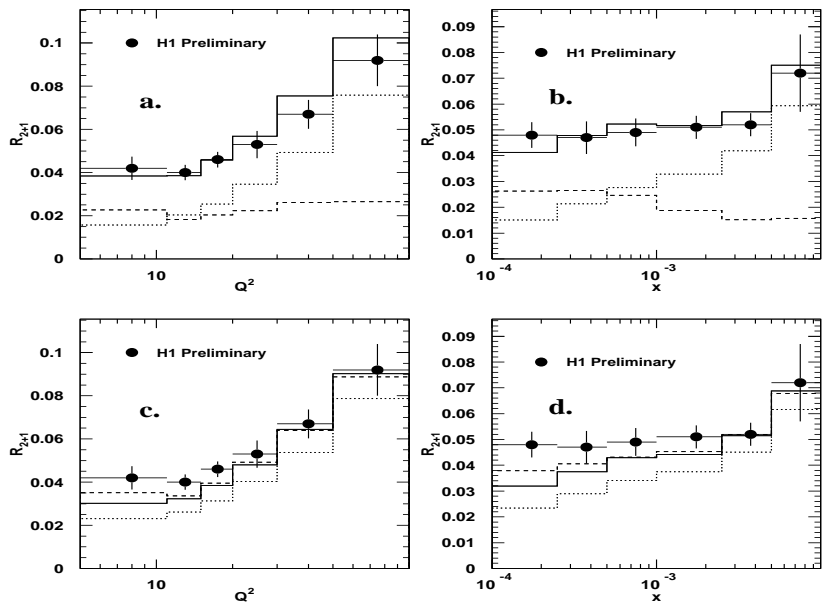
,width=11cm,height=8cm}
\end{center}
\caption{The $2+1$ jet ratio $R_{2+1}$ as a function of
$Q^2$ and $\xbj$. 
The data points
are preliminary H1 data \protect\cite{H1_2+1jets_data}. 
In $2a._)$ and $b._)$ the dotted line is the RAPGAP prediction
for the direct contribution,  the dashed line for resolved photon contribution
and the solid line is the prediction for the sum of direct and
resolved contributions
(using $\mu^2=Q^2 + 4 p_T^2$,  GRV HO for the proton and
SaSgam 2 DIS for the virtual photon).
In $2c._)$ and $d._)$ the solid line is the RAPGAP  prediction
for the sum of direct and resolved contributions 
with $\mu^2=Q^2 + p_T^2$,   
the dashed line with $\mu^2 = 4 p_T^2$ and the dotted line with
$\mu^2 = p_T^2$. 
\label{2+1jet_res+dir}}
\end{figure}
It has been observed~\cite{H1_2+1jets_data}, that  $direct$ 
LO Monte Carlos like RAPGAP  $direct$ and LEPTO  as well as 
NLO calculations are unable to describe the data especially in the 
low $Q^2$ region of $Q^2< 30$ GeV$^2$. However ARIADNE agrees with the 
data remarkably well. In Fig.~\ref{2+1jet_res+dir}$a._)$ and $b._)$ the 
H1 measurement~\cite{H1_2+1jets_data}
of $R_{2+1}$  as a function of $Q^2$ and $\xbj$  is
shown together with the
$direct$ contribution and the sum of 
$direct$ and resolved photons
obtained with RAPGAP. 
The parameter setting is the same as for the forward jet analysis.
We find remarkable agreement (using $\mu^2=Q^2 + 4 p_T^2$)
between the data and the 
Monte Carlo prediction showing that especially in the region of 
phase space where $Q^2< p_T^2$, resolved photons play an important role.
Note that here also the dominant contribution from resolved photon processes 
comes from $qg \to qg$. 
\par
We have checked $R_{2+1}$ as a function of $Q^2$ and $\xbj$ also for 
different scales $\mu^2=Q^2 + p_T^2$, $\mu^2 = 4 p_T^2$ and
$\mu^2 = p_T^2$ 
used  for the resolved photon contribution. The contribution of
resolved photons to $R_{2+1}$ is larger for the largest scale
$\mu^2=Q^2 + 4 p_T^2$,
since there is a larger phase space for DGLAP evolution in the photon.
The results are shown in Fig.~\ref{2+1jet_res+dir}$c._)$ and $d._)$
Again we observe the same trend, that including resolved photons in
DIS,  the description of the measurements is considerably improved for
all different choices of the scales $\mu^2$. 
%%%%%%%%%%%%%%%%%%%%%%%%%%%%%%%%%
\section{Discussion}
%%%%%%%%%%%%%%%%%%%%%%%%%%%%%%%%%
A scale dependence is expected in all leading order calculations, 
but it becomes
more important when the concept of resolved photons is applied to DIS:
in order to resolve any internal structure of the virtual photon, the 
scale involved has to be larger than the virtuality of the photon
$Q^2$:
$ \mu^2 > Q^2 $.
Using a large scale like $\mu^2 = Q^2 + 4 p_T^2$ 
results in a larger magnitude of 
the parton density, both on the photon and proton side, at fixed values of
fractional parton momenta $\xi_i$, 
due to the larger phase space for DGLAP evolution. 
This effect is compensated to some extent by the
smaller value of $\alpha_s$ at the same scale. 
\par
In case of the forward jet cross section the stronger scale dependence 
becomes understandable because of the requirement $0.5 < Q^2/p_T^2 < 2$.
Although here $p_T$ is measured in the lab system this cut restricts
severely the phase space for resolved photons.
\par
With
the concept of resolved photons, 
experimentally observed phenomena in DIS,
 such as the forward jet cross section and
the 2+1 jet rate can be well described,
 provided a reasonable large scale for the hard scattering process
is used. 
This concept and its implementation into a full hadron level 
Monte Carlo generator can be seen as 
 a  model alternative to 
BFKL calculations~\cite{Bartels_fjets}.
The next step would be to search for regions in phase space where
the two descriptions do not agree but the data can only be described by one 
of them.
%%%%%%%%%%%%%%%%%%%%%%%%%%
\section*{Acknowledgment}
%%%%%%%%%%%%%%%%%%%%%%%%%%
I am grateful for many fruitful discussions with L.  J\"onsson and H. K\"uster.
\par
Without the great time and fun I still have with Antje it would not have 
 been possible  to search for free living pomerinos on the Canary Islands.
%
%-----------------------------------------------------------------------
%

\end{document}